\documentclass[10pt, conference, letterpaper]{IEEEtran}

\usepackage[T1]{fontenc}
\usepackage[utf8]{inputenc}

\usepackage[cmex10]{amsmath} 
\usepackage{amssymb,amsfonts}
\usepackage{mathtools}
\usepackage{braket}
\DeclareMathOperator{\Err}{Err}
\DeclareMathOperator{\negl}{negl}
\DeclareMathOperator{\nnegl}{non-negl}

\usepackage{amsthm}
\newtheorem{theorem}{Theorem}

\newtheorem{definition}[theorem]{Definition}

\usepackage{graphicx}
\usepackage{url}
\usepackage{cite}
\usepackage{hyperref}
\usepackage[capitalize]{cleveref}

\begin{document}

\title{Comparison of Quantum PUF models}

\author{
\IEEEauthorblockN{Vladlen Galetsky, Soham Ghosh, Christian Deppe and Roberto Ferrara}
\IEEEauthorblockA{Technical University of Munich, Institute for Communications Engineering,  Munich, Germany\\
Email: \{vladlen.galetsky, soham.ghosh, christian.deppe, roberto.ferrara\}@tum.de}
}
\maketitle

\begin{abstract}
Physical unclonable functions (PUFs) are hardware structures in a physical system (e.g. semiconductor, crystals etc.) that are used to enable unique identification of the semiconductor or to secure keys for cryptographic processes. A PUF thus generates a noisy secret reproducible at runtime. This secret can either be used to authenticate the chip, or it is available as a cryptographic key after removing the noise. Latest advancements in the field of quantum hardware, in some cases claiming to achieve quantum supremacy, highly target the fragility of current RSA type classical cryptosystems. As a solution, one would like to develop Quantum PUFs to mitigate such problem. There are several approaches for this technology. In our work we compare these different approaches and introduce the requirements for QTOKSim, a quantum token based authentication simulator testing its performance on a multi-factor authentication protocol.
\end{abstract}

\begin{IEEEkeywords}
Quantum PUF - Quantum memory - Authentication - Simulator
\end{IEEEkeywords}

\section{Introduction}

While the development and realisation of fifth-generation (5G) mobile communications is in full swing, concrete strategic considerations for the following sixth generation (6G) with a target horizon in 2030 are already beginning in research and industry. Important innovation leaps are expected in 6G with regard to intelligent and environment-adaptable communication, sustainability, availability and security of critical infrastructure. This is where quantum Physical Unclonable Functions (PUFs) may play an important role. So far only, classic PUFs have been practically used. These ideas are to be extended to quantum technologies.

The idea of considering hardware assumptions in designing cryptographic protocols has been a robust field of research since the past years. The idea being first introduced by Katz~\cite{10.1007/978-3-319-56620-7_14} eliminated the need for trusting a designer party or relying on computational assumptions as security came out from the inherent physics of the hardware. One such hardware assumption is the PUF model.  

PUFs are devices that utilise the uncontrollable random disorders which occur during their manufacturing process as a tool for security. These uncontrollable random disorders make them hard to clone for practical purposes. The behaviour of a PUF can be characterised by a set of \textit{Challenge-Response Pairs} (CRP) which are extracted through physically querying the PUF and measuring its responses. Due to this inherent uncontrollable randomness of PUFs it is hard to predict the responses for given challenge. Even the manufacturer who can have access to a set of CRPs cannot predict the response for a new challenge. This property makes PUFs different
from other hardware tokens in the sense that the manufacturer of a hardware token is completely aware of the behaviour of the token they have built~\cite{10.1007/978-3-642-22792-9_4}

There have been a lot of research in the domain of classical PUFs (cPUFs). However, most of the cPUFs are vulnerable to side channel attacks~\cite{10.1007/978-3-030-16350-1_4} and/or machine learning attacks~\cite{10.1007/978-3-662-53140-2_19}. With the advancement of Quantum Technology it is worth studying whether it could boost or threaten the security of cPUFs. In this paper we review the advancements in the field of Quantum PUFs (qPUFs) by doing a comparative study on two major models, namely the Quantum Read-out of PUF or \textit{(QR)-PUF}~\cite{S12} and the unknown Unitary PUF~\cite{ADDK21}. 

Lastly, at the end of this paper we also do a comparative study on existing simulation tools thus sketching a way to implement qPUF based simulations on a new simulator (QTOKSim). We would also leave out error analysis and proof of security in these models. We would rather like to compare these models at the fundamental level of assumptions.

\section{Notation}
We use lowercase bold letters, e.g.\ $\textbf{x}$, for bitstrings and caligraphic letters, e.g.\ $\mathcal{X}$, for the the space of bitstrings, namely $\{0,1\}^n$ for some integer $n$. 
We use uppercase letters, e.g.\ X, for random variables and and we denote vectors as $\vec{x}$.

For pure quantum states we use the Dirac bra-ket notation $\ket{\psi}$ and lowercase greek letters, e.g.\ $\rho$ and $\sigma$ for general density-matrix states.
The Ulhmann fidelity of two quantum states $\rho$ and $\sigma$ is denoted $0 \leq F(\rho,\sigma) \leq 1$.
Finally, we say the quantum states are $\mu$-distinguishable if $F(\rho,\sigma)\leq 1- \mu$ and $\nu$-indistinguishable if $1-\nu\leq F(\rho,\sigma)$.

\begin{figure}
    \centering
    \includegraphics[scale = 0.5]{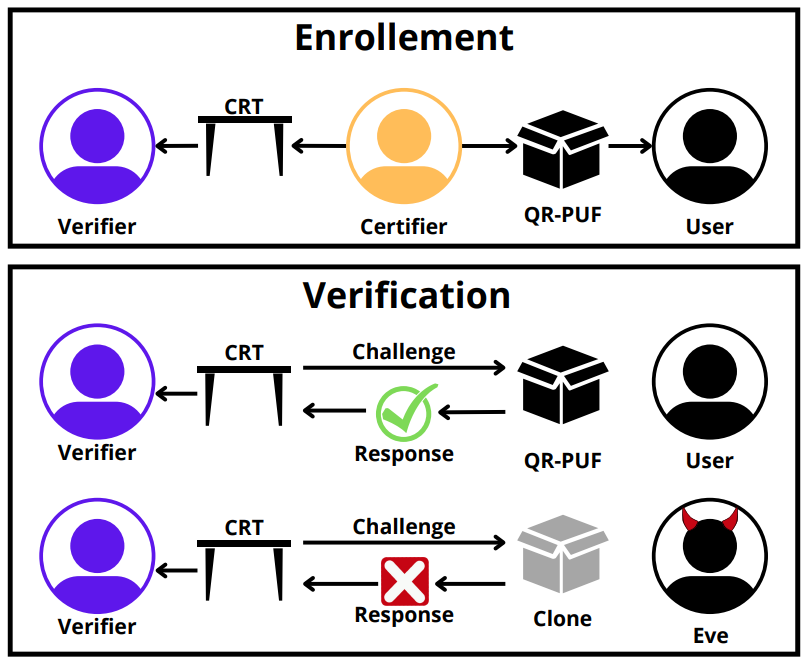}
    \caption{The figure describes the two stages of the authentication scheme: Enrollment and Verification. \textbf{\textit{Enrollment:}} Certifier generates the Challenge-Response Table (CRT) by quering the QR-PUF and hands it over to the Verifier while the QR-PUF is given to the user. \textbf{\textit{Verification:}} The verifier interacts with the user's QR-PUF through a Quantum channel and verifier's his/her/their identity using the CRT. }
    \label{fig:qr-puf}
\end{figure}

\section{(QR)-PUF}

The Quantum Read-out of PUFs~\cite{S12} is one of the very first models for authentication using Quantum PUF. This model essentially exploits the \textit{"no-cloning theorem"} for quantum states as a tool for security. 
The authentication scheme for QR-PUFs is divided into two segments, \textit{enrollment stage} and \textit{verification stage}, see \cref{fig:qr-puf}.

\subsection{Enrollment Stage}

In this section we would rigorously describe the construction of the Challenge-Response Table (CRT). Since, not all states are experimentally implementable or they do not lead to distinguishable responses, the certifier selects $N \leq 2^{n}$ challenges $\textbf{x}_{i} \in \mathcal{X} \subseteq \{0,1\}^{n}$ , where $\mathcal{X}$ is implemented by a set of non-orthogonal states $\{|x_{1}\rangle,\cdots ,|x_{N}\rangle \} \in \mathbb{C}^{2^{\lambda}}$.

The non-orthogonality is very crucial in a sense that we can exploit the power of no-cloning theorem~\cite{Wootters1982} which states that there does not exist an universal quantum cloning machine. Only by co-incidence if the qubit to be cloned is orthogonal to the ancilla on which its information is to be copied one can have a cloning machine which does that. This enhances the security of (QR)-PUFs compared to cPUFs, since an
adversary could gain only a limited amount of information about the challenge and the outcome states. 
We work with separable challenge states $|\psi_{x_{i}}\rangle = \bigotimes^{\lambda}_{k = 1 } |\psi_{x_{ik}}\rangle $. The single-qubit challenge states can be written in terms of some complete orthonormal basis, which we denote as 
\begin{equation}
    |\psi_{x_{ik}}\rangle = \cos \theta_{ik} |0\rangle + e^{i\phi_{ik}} \sin\theta_{ik} |1\rangle,
\end{equation}
where $\theta_{ik} \in [0, \pi]$ and $\phi_{ik} \in [0, 2\pi]$.

The certifier sends all states to the (QR)-PUF, collecting the outcome states. The (QR)-PUF is formalised as a $\lambda$-fold tensor product of single qubit unitary gates $\hat{\Phi} = \bigotimes_{k=1}^{\lambda}\hat{\Phi}_{k} $. Applying the (QR)-PUF on a challenge state we get $|\psi_{y_{i}}\rangle = \bigotimes_{k=1}^{\lambda} |\psi_{y_{ik}}\rangle $, where
\begin{equation}
     |\psi_{y_{ik}}\rangle = \hat{\Phi}_{k} |\psi_{x_{ik}}\rangle.
\end{equation}
For every outcome state, $|\psi_{y_{i}}\rangle = \hat{\Phi}|\psi_{x_{i}}\rangle$, the certifier then also constructs maps called \textit{shifters}, namely, $\hat{\Omega}_{i} = \bigotimes_{k=1}^{\lambda}\hat{\Omega}_{ik}$ which can also be written as tensor product of single-qubit unitary maps. These \textit{shifters} are unitary maps defined as
\begin{equation}
    \hat{\Omega}_{ik} |\psi_{y_{ik}}\rangle = |0\rangle.
\end{equation}
It can be easily proved that the \textit{shifters} are indeed unitary maps. It is possible to implement $\hat{\Omega}_{i}$ for each $\hat{\Phi} |\psi_{x_{i}}\rangle$ because the certifier can fix a challenge state and query the PUF several times and then apply quantum state tomography to characterise each output state. 

Now instead of having to change the single-qubit measurement basis for each response qubit one can measure in only the basis $\{ |0\rangle, |1\rangle \}$ by applying the suitable shifters to $|\psi_{y_{i}}\rangle$.

From the definition of $\hat{\Omega}_{ik}$ for every output qubit we would measure the state $|0\rangle$ and the results of the measurement would form a string of length $\lambda$ consisting of all zeros, for example $\textbf{o}_{i} = \textbf{0} = 00\cdots0$. If there is error (noise from environment, incorrect implementation etc.), then the Hamming weight of $\textbf{o}_{i}$ will give the estimate of it.

The certifier can now parametrize the experimental system that implements the shifters in terms of the number of configurations that it must assume to implement a specific $\hat{\Omega}_{i}$. The set of necessary configurations are taken to be discrete. Thus, a given $\hat{\Omega}_{i}$ can be represented by a classical string $\textbf{w}_\textbf{i} \in \mathcal{W}$ of length $l_{w}$. The length $l_{w}$ depends on the entropy of the shifters and, consequently, on the outcome states (for some implementations, methods to analyze such an entropy have been derived~\cite{EntropyProof1}). 

Finally a classical string $\textbf{y}_{\textbf{i}}$ can be defined which would be called \textit{outcome} string corresponding to the challenge string $\textbf{x}_{\textbf{i}}$ of length $l = l_{w} + \lambda$ as
%
%
%
\begin{equation}
    \textbf{y}_{\textbf{i}} = \textbf{w}_{\textbf{i}} \| \textbf{o}_{\textbf{i}}
\end{equation}
where $\|$ is the concatenation of bit strings.

Finally, the certifier stores the input and output strings and constructs the CRT, $(x_{i},y_{i})$. Then he gives this table to the verifier.

\subsection{Verification Stage}

In the verification stage, the verifier selects a random challenge state, $|\psi_{x_{i}}\rangle$ and sends to the user. The user then interacts it with his (QR)-PUF and sends output state $|\tilde{y}_{i}\rangle$ back to the verifier. The verifier extracts the bit string $\tilde{y}_{i}$ from it just like the certifier did. If $y_{i}$ from the CRT matches the response string $\tilde{y}_{i}$ then the authentication is successful.

\section{Unknown Unitary PUF}

The authors of this model~\cite{ADDK21} studied the (QR)-PUF and came up with their model to address various issues of the (QR)-PUF models. Apart from the problem of Quantum Memory (which is still a domain of research) they have identified and solved various underlying issues of the (QR)-PUFs. To begin with, they have formalised the basic reasonable hardware assumptions/requirements for a Quantum PUF thus providing a general framework for studying them. Previously, for example in the case of (QR)-PUFs, there was no general formal structure for the hardware assumptions of the qPUF.

\subsection{The Formal Structure}

We begin by establishing some basic notations and definitions.
Firstly, to formalise the manufacturing process we define a generation algorithm (QGen) that takes the security parameter of the protocol as input and generates an unique identifier for the qPUF:
\begin{equation}
    \mathrm{QGen(\lambda)} \rightarrow \mathrm{qPUF_{\textbf{id}}},
\end{equation}
where \textbf{id} is the identifier of $\mathrm{qPUF_{\textbf{id}}}$ and $\lambda$ is the security parameter. 

Secondly, we need an evaluation algorithm (QEval) which maps any input quantum state $\rho_\mathrm{in} \in \mathcal{H}^{d_\mathrm{in}}$ to an output quantum state $\rho_\mathrm{out} \in \mathcal{H}^{d_\mathrm{out}}$ where $\mathcal{H}^{d_\mathrm{in}}$ and $\mathcal{H}^{d_\mathrm{out}}$ are the input and output Hilbert spaces with dimension $d_\mathrm{in}$ and $d_\mathrm{out}$ respectively:
\begin{equation}
    \rho_\mathrm{out} \leftarrow \mathrm{QEval(qPUF_{\textbf{id}},\rho_\mathrm{in})}.
\end{equation}
If for now we allow the most general form of Completely Positive Trace Preserving maps for QEval then we have:
\begin{equation}
    \rho_\mathrm{out} = \Lambda_{\textbf{id}(\rho_\mathrm{in})}.
\end{equation}
Then we need a formal definition of a Test algorithm ($\mathcal{T}$) to test the equality between unknown quantum states. 

\begin{definition}[Testing algorithm~\cite{ADDK21}]
Let $\rho^{\otimes k_{1}}$ and $\sigma^{\otimes k_{2}}$ be $k_{1}$ and $k_{2}$ copies of two quantum states $\rho$ and $\sigma$ respectively. A Quantum Testing Algorithm $\mathcal{T}$ is a quantum algorithm that takes as input the tuple $(\rho^{\otimes k_{1}},\sigma^{\otimes k_{2}})$ and accepts $\rho$ and $\sigma$ as equal (outputs probability 1) with the following probability
\begin{align*}
    \Pr[1 \leftarrow \mathcal{T}(\rho^{\otimes k_{1}},\sigma^{\otimes k_{2}})] &= 1 - \Pr[0 \leftarrow \mathcal{T}(\rho^{\otimes k_{1}},\sigma^{\otimes k_{2}})] \\ &= f(k_{1},k_{2},F(\rho,\sigma)),
\end{align*}
where $F(\rho,\sigma)$ is the fidelity of the two states with $f$ following the limits $ \forall (k_{1},k_{2})$ 
\begin{equation}
    \begin{aligned}
     \lim_{F(\rho,\sigma) \rightarrow 1} f(k_{1},k_{2},F(\rho,\sigma)) &= 1 - \Err(k_{1},k_{2}),  \\
     \lim_{k_{1},k_{2} \rightarrow \infty } f(k_{1},k_{2},F(\rho,\sigma)) &= F(\rho,\sigma), \\
     \lim_{F(\rho,\sigma) \rightarrow 0} f(k_{1},k_{2},F(\rho,\sigma))   &= \Err(k_{1},k_{2}),
    \end{aligned}
\end{equation}
where $\Err(k_{1},k_{2}) > 0 $ characterises the error of the test algorithm and $F(\rho,\sigma)$ the fidelity of the states. 
\end{definition}

We can formally define a qPUF as follows.

\begin{definition} [Quantum Physical Unclonable Functions~\cite{ADDK21}]
Let $\lambda$ be the security parameter and $\delta_{r},\delta_{u},\delta_{c} \in [0,1]$ be the robustness, uniqueness and collision-resistance thresholds. A ($\lambda,\delta_{r},\delta_{u},\delta_{c}$)-qPUF includes the algorithms: QGen, QEval $\mathcal{T}$ and satisfies the following requirements: 
\begin{itemize}
    \item \emph{$\delta_{r}$-Robustness}: For any qPUF generated by QGen($\lambda$) and evaluated using QEval on any two input states $\rho_\mathrm{in}$ and $\sigma_\mathrm{in}$ that are $\delta_{r}$-indistinguishable, the corresponding output quantum states $\rho_\mathrm{out}$ and $\sigma_\mathrm{out}$ are also $\delta_{r}$-indistinguishable with overwhelming probability,
    \begin{equation}
        \Pr[\delta_{r}\leq F(\rho_\mathrm{out},\sigma_\mathrm{out} ) \leq 1] = 1 - \negl(\lambda)
    \end{equation}

    \item \emph{$\delta_{u}$-Uniqueness}: For any two qPUFs generated by the QGen algorithm, i.e. $qPUF_{\mathbf{id}_{\mathbf{i}}}$ and $qPUF_{\mathbf{id}_{\mathbf{j}}}$, the corresponding CPT map models $\Lambda_{\mathbf{id}_{\mathbf{i}}}$ and $\Lambda_{\mathbf{id}_{\mathbf{j}}}$ are $\delta_{u}$-distinguishable with overwhelming probability,
    \begin{equation}
        \Pr[\|(\Lambda_{\mathbf{id}_{\mathbf{i}}} - \Lambda_{\mathbf{id}_{\mathbf{j}}} )_{i \neq j} \|_{\diamond} \geq \delta_{u} ] = 1 - \negl(\lambda)
    \end{equation}
    
    \item \emph{$\delta_{c}$-Collision-Resistance}: For any $\mathrm{qPUF_{\mathbf{id}}}$ generated by QGen($\lambda$) and evaluated by QEval on any two input states $\rho_\mathrm{in}$ and $\sigma_\mathrm{in}$ that are $\delta_{c}$-distinguishable, the corresponding output states $\rho_\mathrm{out}$ and $\sigma_\mathrm{out}$ are also $\delta_{c}$-distinguishable with overwhelming probability,
    \begin{equation}
         \Pr[0\leq F(\rho_\mathrm{out},\sigma_\mathrm{out} ) \leq 1- \delta_{c}] = 1 - \negl(\lambda).
    \end{equation}
\end{itemize}
\end{definition} 
Then it is proven in the following theorem that if the input and output Hilbert spaces for the qPUF have same dimension then the QEval are arbitrarily close to unitary maps instead of being CPT maps. Thus, we work in this regime where the qPUF can be modelled with unitary maps instead of CPT maps. This is shown by the following theorem:

\begin{theorem}[{\cite{ADDK21}}]
Let $\mathcal{E}(\rho)$ be a CPT map as follows:
\begin{equation}
    \mathcal{E}(\rho) = (1- \epsilon)U\rho U^{\dagger} + \epsilon \tilde{\mathcal{E}}(\rho)
\end{equation}
where U is an unitary transformation, $\tilde{\mathcal{E}}$ is an arbitrary (non-negligibly) contractive channel and $0\leq \epsilon \leq 1$. Then $\mathcal{E}(\rho)$ is a $(\lambda, \delta_r,\delta_c)$-qPUF for any $\lambda$, $\delta_{r}$ and $\delta_{c}$ with the same dimension of domain and range of Hilbert space if and only if $\epsilon = \negl(\lambda)$. 
\end{theorem}

Then we move on to the key part of this model which is the notion of \emph{Unknown Unitary} transformation that we impose to enhance the security. This idea makes this model much more reasonable and secure than the (QR)-Puf model as well. 

\begin{definition}[Definition of Unknown Unitary Transformation~\cite{ADDK21}]
We say that an unitary transformation $U^{i}$, over a D-dimensional Hilbert space $\mathcal{D}^{D}$ is called Unknown Unitary Transformation, if for all Quantum-Polynomial-Time adversaries $\mathcal{A}$ the probability of estimating the output of $U^{u}$ (where superscript u is a uniformly distributed classical random variable) on any randomly picked state $|\psi \rangle \in \mathcal{H}^{D}$ is at most negligibly higher than the probability of estimating the output of a Haar random unitary operator on that state
\begin{align}
    \Pr_{U^{i}} [F(\mathcal{A}(|\psi\rangle), U|\psi\rangle )\geq \nnegl(\lambda)] &- \notag  \\ 
    \Pr_{U^{\mu}} [F(\mathcal{A}(|\psi\rangle), U^{\mu}|\psi\rangle )\geq \nnegl(\lambda)] &= \negl(\lambda)
\end{align}
where $\mu$ is the Haar measure (uniformly distributed classical random variable) and the average probability is taken over all possible states $|\psi\rangle$.
\end{definition}

\subsection{Authentication scheme}
The authentication scheme is similar to that of (QR)-PUF, only difference is that in this model we have a Quantum CRT. The PUF is queried with quantum states and the physical quantum states are stored in a quantum memory. Then the verification is done using the "Testing Algorithm" which checks if the response state is the same as the one stored in quantum memory.

\section{Brief Comparison}

There are several issues with the (QR)-PUF model that are not there with the unknown unitary PUF model. 
\begin{itemize}
    \item The CRT consists of essentially classical strings, thus we did not quite leverage the power of quantum states.
    \item The unitary of the PUF is completely known, hence an adversary with an efficient Quantum Computer can just simulate the unitary to predict the response states.
    \item There is a trade off between noise and unclonability in the (QR)-PUF model.
d    \item The certifier knows the CRT table which becomes a trusted party.
\end{itemize}
The Unknown Unitary model takes care of all these issues but however still requires quantum memory to perform. 
For future works, one could consider fewer hardware assumptions or could try to model PUFs with general non-unitary quantum channels other than unitary.

\section{Simulation}

\subsection{Simulator implementation}
To study even further current token based architectures, from the DIQTOK, HybridQToken, NEQSIS, Q-ToRX, QPIS and QuaMToMe projects, a robust simulator is proposed in this section. The main differences in the hardware implementation of each project aims mainly to extend the storage time of quantum information, being this a crucial variable not only for token based authentication but also for the whole field of quantum computation. The implementations range from nitrogen vacancy centers in diamond (DIQTOK) with alternative combination of photonic integrated circuits (QPIS), full classical implementation into a microchip of a quantum memory (HybridQToken), rare earth ions (NEQSIS) and electron/nuclear spin ensembles as storage units (QuaMToMe), to gas cells with xenon and rubidium (Q-ToRX).

The requirements for the QTOKSim simulator are then described as follows:
\begin{itemize}
  \item Based on the theory of each project, an Hamiltonian structure shall be proposed depending on the memory performance.
  
  \item For each project, using deep quantum machine learning, a behavioral description of each hardware component shall be defined. As a result, this would allow to define a realistic Hamiltonian for each system.
  
  \item The quantum circuit responsible for the authentication protocol shall be corrected for each interaction with a gradient descent optimizer, by comparison between the theoretical and realistic Hamiltonians.
  \item  A realistic implementation for quantum PUF's shall be considered, based on the performance of other systems. 
  \item A number of attacks shall be considered from man in the middle to combined attacks. This way, the simulator would allow to study the fragility of each system, proposing new solutions to prevent it. 
\end{itemize}

Before introducing the architecture of such simulator, we first need look into the available proposals within the field.

\subsection{State of the art quantum simulators}
To the knowledge of the authors, no simulator is know to solely work realistically with quantum authentication protocols. Such system should be distinctly divided into two fields: the encoding and authentication of the secret key via a quantum circuit and quantum network to distribute such information.

A wide range of simulators are available for quantum networks such as Netsquid~\cite{1k}, Sequence~\cite{1l}, QuisP~\cite{2a}, QuNetSim~\cite{2b} and SQUANCH~\cite{2c}. 
They differ in terms of the architecture used, for example Netsquid uses an hardware library, allowing users to describe diamond colour center memories in order to characterize realistically the decoherence $\tau_1$ and $\tau_2$ parameters. Scalability to more nodes, time optimization method, universal quantum computation capability and the respective time-event engine (discretization) are some distinctive parameters between each simulator.

\begin{figure*}
\centering
\includegraphics[width=1.0\textwidth,clip]{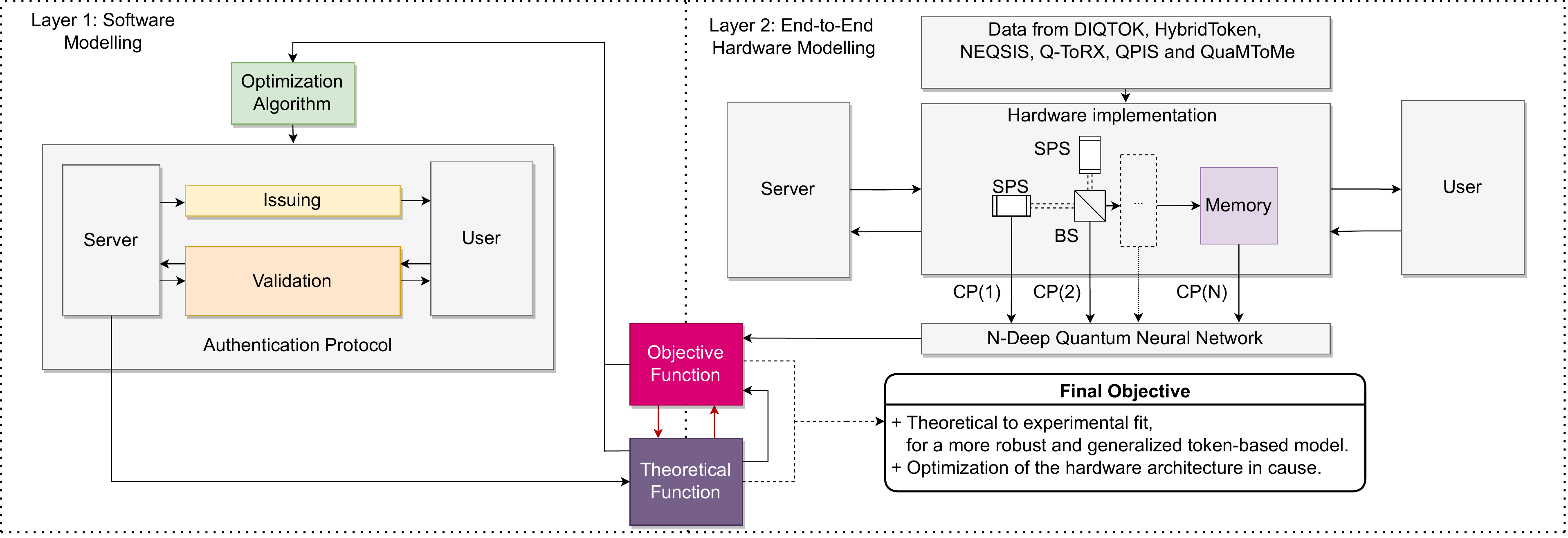}
\caption{
QTOKSim architecture based on a theoretical (I) and physical (II) layers. An end-to-end configuration is described for hardware correction and circuit optimization.}
\label{sim1}       
\end{figure*}

In terms of the construction of the circuits, there are mainly IBM Qiskit~\cite{1h}, Intel Quantum Simulator~\cite{1i}, LIQUi$\ket{}$~\cite{lol} and Pennylane~\cite{1j}. Qiskit provides with a framework for a wide range of quantum circuits and algorithm implementations either classically, with fake or realistic backends going all the way from the Falcon r5.11H~\cite{1g} to Eagle r1~\cite{1f} quantum processors. Respectively, those backends describe the system realistically by mainly characterizing the realistic gate error ($g_{err}$) probabilities of each basis gate on each qubit. Gate lengths of each basis gate on each qubit, the $\tau_1$ and $\tau_2$ decoherence of each qubit and the readout error probability of each qubit.

On the other hand, Pennylane differentiates from other simulators by focusing more on quantum network and optimization algorithms with hybrid quantum and classical models. The advantage of such simulator is that it is device independent and easily complements with previous methods.  

\subsection{Architecture of QTOKSim}
Based on the previous analysis of the state of the art, the following structure is suggested for the simulator:
\begin{enumerate}
  \item The network behaviour network is based on the Netsquid simulator performance.
  \item The quantum circuit is computed using Qiskit, by exploiting the realistic error performance from the hardware. 
  \item A NISQ type algorithm is the basis of the structure, where a corrective method between the theory and hardware simulation is introduced. 
  \item To perform quantum authentication attacks and to describe the hardware Hamiltonian, Pennylane package will be used to compute the respective quantum emulators and quantum neural network algorithms.
\end{enumerate}

The architecture is described in Fig.\ref{sim1} with two main layers, a theoretical one incorporating the quantum token authentication protocol, quantum circuit, the nodes and the network composed by the issuing and validation tasks. The second layer works as a realistic backend allowing to perform an end-to-end optimization on the structure. Generally, we can describe the quantum token circuit by a combination of a state encoder, memory and basis change decoder.

\subsection{Quantum Multi-factor Authentication scheme}
\label{qm}
A behaviour analysis of layer one was studied by implementing a quantum multi-factor authentication scheme from the works of Hazel~\cite{1e} and Dmitry G.\cite{1d} . The main idea of this scheme consists on the Hidden Matching Problem ($\mathit{HMP}_{4}$) for four states and can be described as follows:

\begin{enumerate}
    \item The protocol has an issuing and a validation phase.

    \item Issuing phase: the protocol follows $\mathit{HMP}_{4}$ for which $\ket{\alpha}=\frac{1}{2}\sum_{1\leq i \leq4}(-1)^{x_{i}}\ket{i}$, where $x \in \{0, 1\}^{4}$ this allows the server to encode 4-k strings of information into 2-qubit registers, for which the user stores that set a result. A classical Id-token is also provided via a public channel for a first user identification.
    
    \item Validation phase: the user and the server agree on a subset of unused 2-qubit quantum registers. This is done by the server choosing a set $L_{s}$ of t indices randomly and the user generating a subset $L_{d}\subset L_{s}$ of size $\frac{2t}{3}$. For each chosen qubit the server also randomly chooses a set of basis ($m_i$ , $i = 0, 1$) for measurement. 
    \item The user measures the quantum registers corresponding to the elements from the basis set. The corresponding results are encoded classically into $\frac{2t}{3}$ pairs ($a_i$, $b_i$), such that $(x_i, m_i, a_i, b_i)$ $\in$ $\mathit{HMP}_4$ for all $i \in L_{s}$, as seen in the work of Dmitry~\cite{1d}. Via a public channel the user sends back to the server the ($a_i$, $b_i$) pairs. Following the $\mathit{HMP}_4$ condition: $m, a, b \in \{0, 1\}$, we can demonstrate that $(x, m, a, b)$ $\in$ $\mathit{HMP}_4$. This condition allows the server to verify the user using the following relation~\cite{1d}: 
    \[
    b = 
    \begin{dcases}
        x_{1} \oplus x_{2+m},& \text{if } a = 0\\
        x_{3-m} \oplus x_{4},& \text{if } a = 1
    \end{dcases}
    \]
\end{enumerate}

By using Qiskit for the circuit formation and Netsquid for the network, the simulation of such system was performed by defining eight entangled and other eight non-entangled independent states to encode 4 bits of information quantumly.

\begin{figure}
\centering
\includegraphics[scale = 0.35]{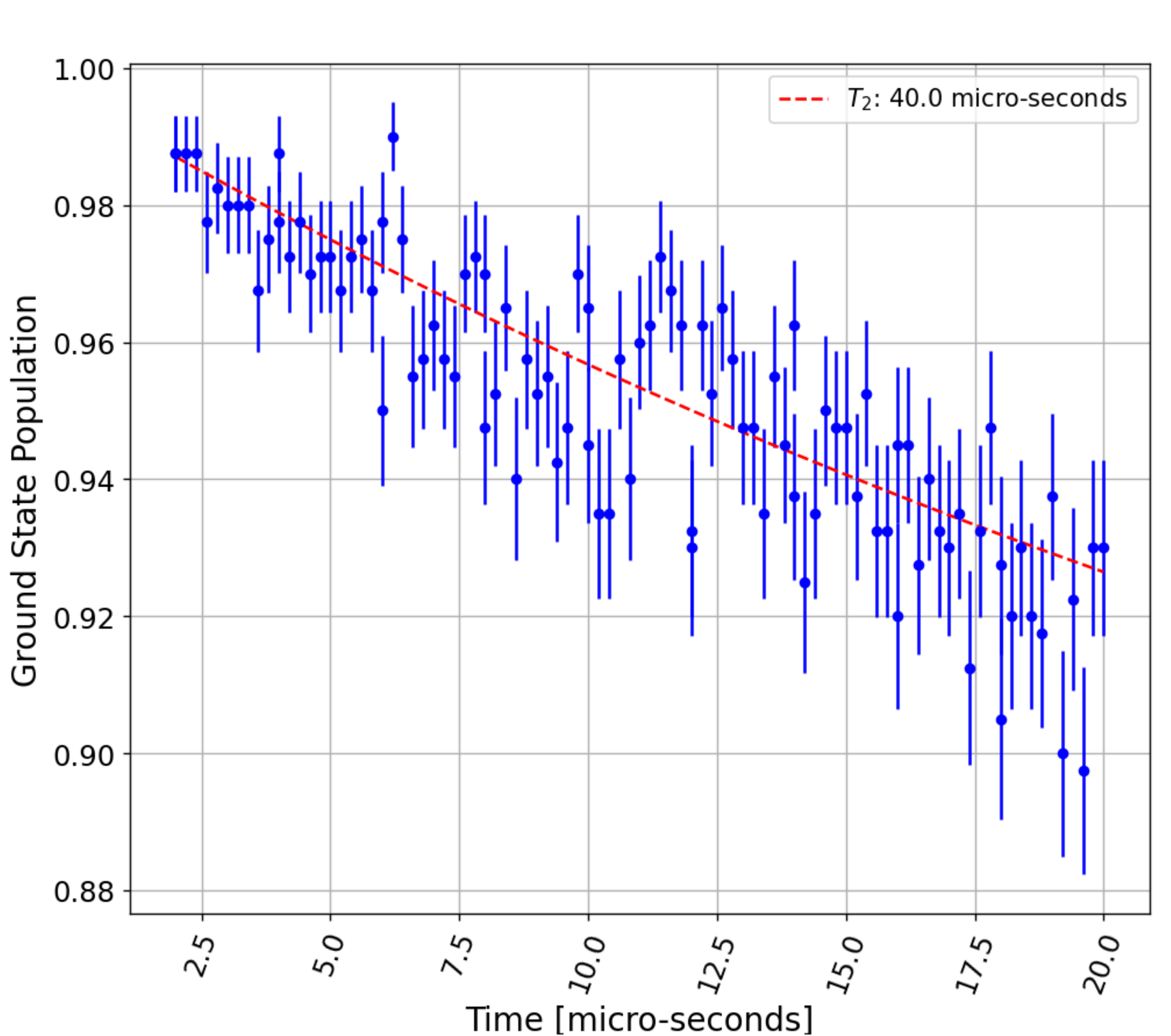}
\caption{
Quantum decoherence in the user's memory simulated with Qiskit IBM Oslo backend. }
\label{sim2}       
\end{figure}

A IBM Oslo Falcon r5.11H processor~\cite{lula} was used which realistically considers all sources of error on a real quantum device. The degree of dephasing  for the user's memory in this model is of the type: $f(x)=Ae^{-\frac{t}{T_{2}}}+B$. The system specifications such as the average readout error:
$1.581\times10^{-2}$ and the average error induced by having the qubit idle for a typical gate time: $3.654\times10^{-4}$ are presented in~\cite{lula}.
The results for the quantum memory dephasing can be viewed in \cref{sim2}, where for example for a pass-time of $10\mu s$ a $(4.4\pm4.50)\%$ probability is obtained to pass from a state $\ket{+}$ to $\ket{-}$.

Considering that quantum memories are a central point for PUF and quantum token authentication protocol implementation, a more in-depth analysis on this topic should be performed. Just by considering different architectures the performance may vary: by looking at the work of Pengfei Wang et al.\cite{1b} the decoherence rate reaches an order ($\mathcal{O}$) of minutes ($\mathcal{O}(8)$) for the same population concentration. However, by looking at the work of Edwar Xie~\cite{1a} the performance is actually worse by an order of $\mathcal{O}(1)$.  

\section{Conclusions}
In this work we introduce the Q.TOK project and review different models of quantum PUF's as well as their advantages or disadvantages in a theoretical framework for security implementation.

Considering the necessity not only to compare the performance of quantum PUF's between theory and reality, but also to compare different quantum token based authentication protocol architectures a new simulator is proposed (QTOKSim). As a central point of the hardware performance is the quantum memory, a dephasing time study is shown in \cref{sim2} reaching an order of micro-seconds for the Quantum Multi-factor Authentication scheme presented in \cref{qm}.

\section*{Acknowledgement}
Christian Deppe and Roberto Ferrara were supported by the German Federal Ministry of Education and Research (BMBF), Grant 16KIS1005.
Christian Deppe acknowledge the financial support by the Federal Ministry of Education and Research of Germany in the programme of ''Souver\"an. Digital. Vernetzt''. Joint project 6G-life, project identification number: 16KISK002.
Vladlen Galetsky, Soham Ghosh and Christian Deppe were supported by the BMBF Project 16KISQ038.
Christian Deppe and Roberto Ferrara are supported by the Munich Quantum Valley, which is supported by the Bavarian state government with funds from the Hightech Agenda Bayern Plus.
\bibliographystyle{IEEEtran}
\bibliography{references}

\end{document}